\documentclass[a4paper,10pt,fleqn]{article}
\usepackage{graphicx}
\usepackage{breakurl}
\usepackage{times}
\usepackage{mathrsfs}
\usepackage{color}
\usepackage{times}
\usepackage[authoryear]{natbib}
\usepackage{mathrsfs}
\usepackage{pdflscape}
\usepackage{array}
\usepackage{amsmath}
\usepackage[normalem]{ulem}
\usepackage{hyperref}
\usepackage{comment}

\begin{document}
\title{Clinical trials impacted by the COVID-19 pandemic: Adaptive designs to the rescue?}

\author{Cornelia Ursula Kunz$^{1}$\footnote{Authors contributed equally}, Silke J\"orgens$^{2}$\footnotemark[1], Frank Bretz$^3$, Nigel Stallard$^4$, \\
Kelly Van Lancker$^5$, Dong Xi$^6$, Sarah Zohar$^7$, Christoph Gerlinger$^{8,9}$\footnotemark[1], Tim Friede$^{10,11}$\footnotemark[1] \footnote{Corresponding author: {\sf{e-mail: tim.friede@med.uni-goettingen.de}}, Phone: +49-551-39-4991, Fax: +49-551-39-4995} \\ \\
\footnotesize $^1$Boehringer Ingelheim Pharma GmbH \& Co. KG, Biberach, Germany\\
\footnotesize $^2$Janssen-Cilag GmbH, Neuss, Germany\\
\footnotesize $^3$Novartis Pharma AG, Basel, Switzerland\\
\footnotesize $^4$Division of Health Sciences, Warwick Medical School, The University of Warwick, Coventry, UK\\
\footnotesize $^5$Department of Applied Mathematics, Computer Science and Statistics, Ghent University, Ghent, Belgium \\
\footnotesize $^6$Novartis Pharmaceuticals, East Hanover, New Jersey, USA \\
\footnotesize $^7$INSERM, Centre de Recherche des Cordeliers, Sorbonne Université, Université de Paris, Paris, France \\
\footnotesize $^8$Statistics and Data Insights, Bayer AG, Berlin, Germany \\
\footnotesize $^9$Department of Gynecology, Obstetrics and Reproductive Medicine, University Medical School of Saarland, Homburg/Saar, Germany \\
\footnotesize $^{10}$Department of Medical Statistics, University Medical Center G\"ottingen, G\"ottingen, Germany\\
\footnotesize $^{11}$DZHK (German Center for Cardiovascular Research), partner site G\"ottingen, G\"ottingen, Germany}

\date{}

\maketitle

\begin{abstract}
Very recently the new pathogen severe acute respiratory syndrome coronavirus 2 (SARS-CoV-2) was identified and the coronavirus disease 2019 (COVID-19) declared a pandemic by the World Health Organization. The pandemic has a number of consequences for the ongoing clinical trials in non-COVID-19 conditions. Motivated by four currently ongoing clinical trials in a variety of disease areas we illustrate the challenges faced by the pandemic and sketch out possible solutions including adaptive designs. Guidance is provided on (i) where blinded adaptations can help; (ii) how to achieve type I error rate control, if required; (iii) how to deal with potential treatment effect heterogeneity; (iv) how to utilize early read-outs; and (v) how to utilize Bayesian techniques. In more detail approaches to resizing a trial affected by the pandemic are developed including considerations to stop a trial early, the use of group-sequential designs or sample size adjustment. All methods considered are implemented in a freely available R shiny app. Furthermore, regulatory and operational issues including the role of data monitoring committees are discussed. 
\end{abstract}

\section{Introduction}
\noindent
In Wuhan, China pneumonia cases of a new pathogen, which was subsequently named severe acute respiratory syndrome coronavirus 2 (SARS-CoV-2), were identified in December 2019 \citep{GuanEtAl2020}. In the meanwhile, the coronavirus diseases 2019 (COVID-19) was declared a pandemic by the World Health Organization (WHO). At the time of writing (end of May 2020), more than 5 million cases were confirmed worldwide according to the COVID-19 Dashboard by the Center for Systems Science and Engineering at Johns Hopkins University (\url{https://coronavirus.jhu.edu/map.html}). To fight the COVID-19 pandemic, a number of clinical trials were initiated or are in planning to investigate novel therapies, diagnostics and vaccines.
Some of these make use of novel, efficient trial designs including platform trials and adaptive group-sequential designs. An overview and recommendations are provided by \citet{StallardEtAl2020}.

While considerable efforts have been made to set up trials in COVID-19, the vast majority of ongoing trials continues to be in other disease areas. In order to effectively protect patient safety in these trials during the COVID-19 pandemic, across the world, clinical trials answering important healthcare questions were stopped, or temporarily paused to possibly re-start later, some with important modifications. Here we consider the impact of the COVID-19 pandemic on running trials in non-COVID-19 indications. The challenges to these trials posed by the pandemic can take various forms including the following: (1) The (amount) of missing data may preclude definite conclusions to be drawn with the original sample size. (2) Incomplete follow-up (possibly not at random) may invalidate the planned analyses. (3) Reduced on-site data monitoring may cast doubt on data quality and integrity. (4) Missed treatments due to the interruptions, but also due to acquiring the SARS-CoV-2 virus may not be random and require a different approach than based on the intention-to-treat principle. (5) Circumstances (in e.g. usual care, trial operations, drug manufacturing) before, during and after the pandemic induced interruptions may differ substantially with impact on interpretability of the clinical trial data, through which the original research question is more difficult or even impossible to answer. (6) Heterogeneity in patients included in the trial associated with the pandemic may impact results. (7) Potential heterogeneity in included patients for multi-center trials, as the prevalence/incidence of infected patients various from region to region. 

\begin{figure}
\centering
\includegraphics[width=0.99\textwidth, angle = 0]{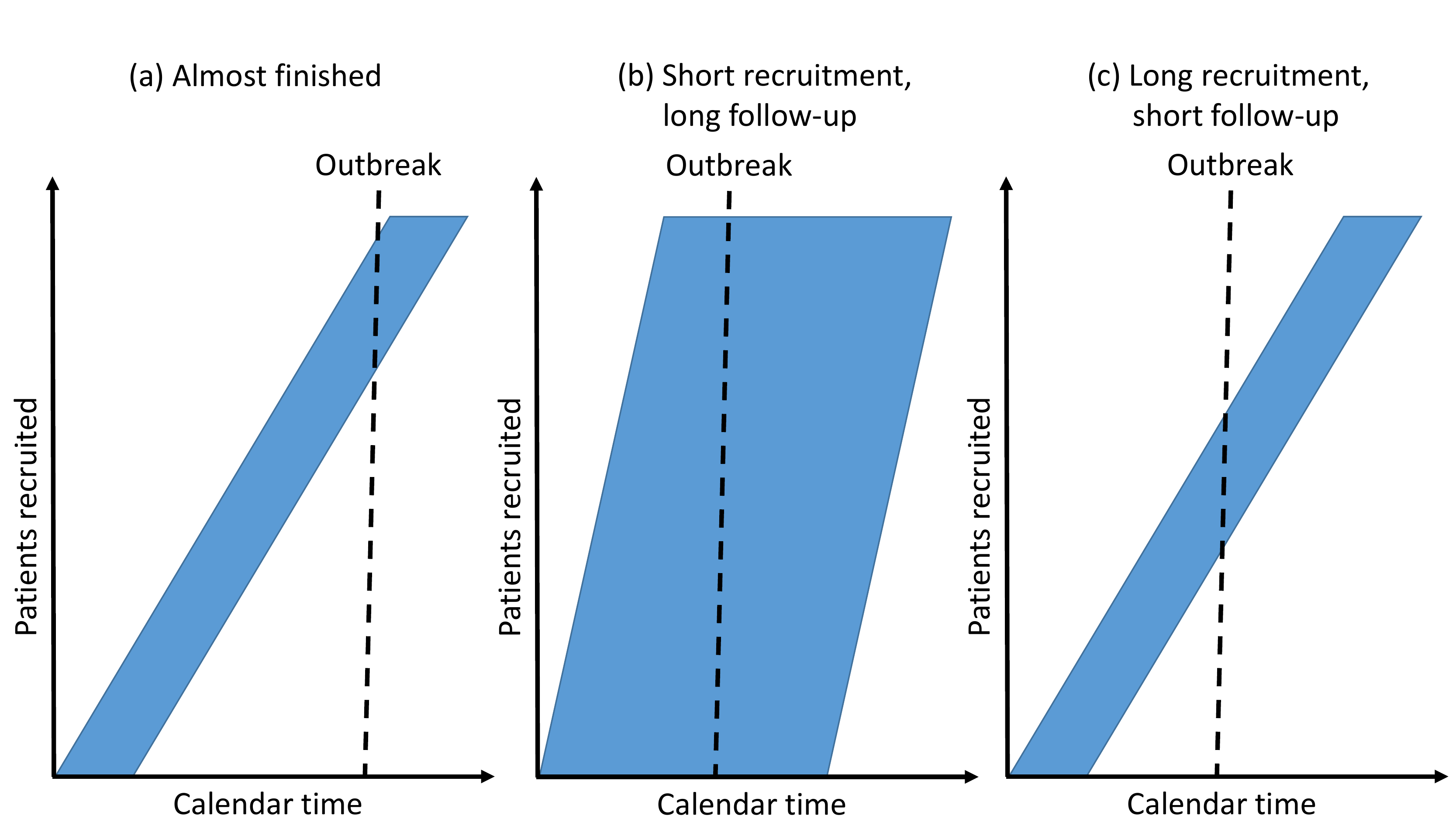}
\caption{Illustration of how the COVID-19 pandemic impacts clinical trials depending on accrual and follow-up.}
\label{fig:intro}
\end{figure}

Regulatory authorities have produced guidance on implications of COVID-19 on methodological aspects of ongoing clinical trials \citep{EMA2020a, EMA2020b, FDA2020}. The EMA guideline states that the current situation should not automatically encourage unplanned interim or early analyses \citep{EMA2020b}. Despite strong scientific reasons to conduct trials as planned, there may be situations where an unplanned or early analysis may be required to minimize the effect of COVID-19 on the interpretability of the data and results. Potential situations include trials where data collection is nearly finished, an interim analysis is planned in the near future, recruitment of new patients is slowing down or interrupted. In particular, the impact of the pandemic depends on the timing of the pandemic compared to the timeline of the trial, the length of follow-up to observe the primary endpoint and the recruitment rate. Figure \ref{fig:intro} illustrates the different scenarios.

For example, when recruitment has been paused and will be restarted after the pandemic, the trial duration will be prolonged. A two-stage adaptive design might then be considered for the clinical trial. An interim analysis evaluating the first stage data, which include participants not affected by COVID-19, should guide the investigators to decide whether it is worthwhile to restart recruitment after the pandemic and with which sample size. Nevertheless, as any unplanned interim analysis needs to protect the trial integrity (e.g., blinding) and validity (e.g., type I error rate) appropriate statistical methodology for testing and estimation at the end of the trial is an essential aspect. The adaptive design literature offers potential solutions to deal with the concerns in modified trial designs. This has also been recognized by \citet{Anker2020} in the context of clinical trials in heart failure, a chronic condition.

The manuscript is organized as follow. In Section \ref{sec:examples} four ongoing clinical trials are introduced which are all impacted by the COVID-19 pandemic. These serve as examples and illustrate the many ways trials might be affected by the pandemic. In Section \ref{sec:general} general comments are made on how adaptive designs might be used to overcome the various challenges posed by the pandemic before the issue of resizing trials is considered in more detail in Section \ref{sec:resizing}. Regulatory and operational issues including the role of data monitoring committees or data safety monitoring boards are considered in Section \ref{sec:regulatory}. In Section \ref{sec:discussion}, we close with a brief discussion.

\section{Motivating examples} \label{sec:examples}
Clinical trials are affected in many different ways by the COVID-19 pandemic. On the one hand, patients may get infected leading to missed visits, missing data, or even COVID-19 related adverse events. On the other hand, the various lockdown and quarantine measures may disrupt the trial conduct: Patients may be unable to attend their scheduled visits or the study medication cannot be delivered to the patients as planned. While these issues apply to all trials recruiting patients or collecting data during the pandemic, they are affected quite differently depending on the stage the trial was in and also depending on the endpoint of the trial.

One important point is still open at the time this paper was written: When and how to restart trials that have had their recruitment interrupted or even the study treatment stopped by the onset of the pandemic? The only thing that seems clear is that the conditions under which a trial is restarted will be very trial specific and can be elaborated only provisionally at the end of this paper.

\subsection{Long acting reversible contraception: The devil is in the detail} \label{sec:contra}
For our first example, consider a study to assess the contraceptive efficacy beyond 5 years up to 8 years of a hormone releasing intrauterine device (IUD) \citep{Jensen2020}. At the onset of the pandemic all participating women had their IUD in place for more than 6 years but only a few had already completed 8 years of treatment. The primary outcome of the trial is the contraceptive failure rate in years 6 to 8 measured by the Pearl Index \citep{Gerlinger2003}. The trial uses a treatment policy estimand, albeit the term \textit{estimand} was not yet invented when the contraceptive trial was conceived. 

COVID-19 related intercurrent events such as missed or postponed visits to the study center can be ignored for the primary analysis. There will be no interruption of study treatments as the IUD has been in the woman’s uterus for 5 years at the beginning of the trial and remains there for up to 8 years in total. Even if the pandemic will last past the scheduled end of the trial, the primary outcome (pregnant yes/no) can still be ascertained even if a woman is not able to attend the final visit in person on time, albeit that according to the statistical analysis plan the continued exposure to the IUD needs to be confirmed by the investigator. Nevertheless, the contraceptive failure rate observed over the whole trial may be impacted not only by a potential loss in confirmed exposure time but also by other COVID-19 related intercurrent events. For instance, a couple who usually commutes long-distance on weekends is not at risk of contraceptive failure during the lockdown if they observe the lockdown living apart, but they are possibly at a higher risk if they observe the lockdown living together. However, given the treatment policy estimand and the very low rate of contraceptive failure with an IUD  \citep{Mansour2010}  these intercurrent events are not likely to be relevant for the interpretation of the trial’s results.

It should be noted that other endpoints of the trial may also be impacted by COVID-19 related intercurrent events. The regular safety assessments planned at the scheduled visits might be at least partially missing if women need to skip the physical visit. While details of adverse events can be obtained by phone, laboratory values will be definitely missing in such instance.

\subsection{The START:REACTS trial: Change in endpoints due to difficulties in recruitment} \label{sec:startreacts}
Our second example is the Subacromial spacers for Tears Affecting Rotator cuff Tendons: a Randomised, Efficient, Adaptive Clinical Trial in Surgery (START:REACTS),  an adaptive design multi-center randomized controlled trial conducted in the United Kingdom comparing arthroscopic debridement with the InSpace balloon (Stryker, USA) to arthroscopic debridement alone for people with a symptomatic irreparable rotator cuff tear \citep{Metcalfe2020}. Recruitment to the trial started in February 2018, with a planned sample size of 221 with the potential to stop the study for efficacy or futility at a number of interim analyses. The primary endpoint was shoulder function 12 months after surgery measured using the Constant Shoulder Score (CS) recorded at a hospital out-patient visit, with assessments taken at 3 and 6 months following surgery also used for interim decision-making \citep{Parsons2019}. 

Due to the coronavirus pandemic, recruitment to the study was delayed by the cancellation of elective surgery in UK hospitals.  The study team are working closely with the Data Monitoring Committee in reviewing the planned timing of the interim analyses to reflect this, and the resulting change in the anticipated numbers of patients with 3, 6 and 12 month follow-up data at different time-points in the study.  The pandemic also threatened disruption of the collection of follow-up data for patients for whom surgery had already been completed, as even prior to lockdown, many patients in the study, a large proportion of whom are in vulnerable groups, were unwilling to attend planned appointments for assessment.  In order to be able to obtain follow-up data from as many patients as possible, the study team decided to change the primary endpoint to be the 12 month measurement of the Oxford Shoulder Score (OSS), as this does not require face-to-face data collection, but can be completed by post or over phone (or app).  As this had originally been included as a secondary endpoint in the study, data were available for all completed patients. The OSS is known to be well correlated to the CS, with the same minimum clinically important difference on a standardized scale, so that the power of the trial is maintained and, as the change was made prior to interim data being observed, there is no loss of trial integrity.

\subsection{The ATALANTE 1 trial: Premature study discontinuation not to endangering sensitive patients during the COVID-19 pandemic} \label{sec:atalante}
The ATALANTE 1 clinical trial (NCT02654587) aimed at evaluating and comparing the medicinal product tedopi (OSE2101) to standard treatment (docetaxel or pemetrexed) as second and third line therapy in HLA-A2 positive patients with advanced NSCLC after failure of immune checkpoint inhibitor. This clinical trial was planned in two stages (1) randomized controlled trial (RCT) on a small sample of patients estimating overall survival rate at 12 months (with about 100 patients) and (2) a RCT comparing overall survival (with about 363 patients in total). After the first stage, 99 patients were included (63 in the experimental arm and 36 in the standard arm), the overall survival rate at 12 months was 46\% (95\% confidence interval: 33\% - 59\%) in the experimental arm and 36\% (95\% confidence interval: 21\% - 54\%) in the standard arm \citep{oseApril1}. The second stage of the study was supposed to include patients during 2020. However, this trial was stopped because of the COVID-19 pandemic. Indeed, as patients were suffering from lung cancer, the DSMB decided that it was too risky to continue. They stated that it was impossible to expose patients suffering from lung cancer to COVID-19 infection, this could endanger them and may end up biasing the results of the trial \citep{oseApril4}. As the results of the first stage were promising, the trial stakeholders decided to shortly discuss with the FDA and the EMA asking whether an additional clinical trial would be required, knowing that there are crucial treatment needs in this indication.

\subsection{The CAPE-Covid and the CAPE-Cod (Community-Acquired Pneumonia: Evaluation of Corticosteroids) studies: Embedding a COVID-19 trial within an ongoing trial} \label{sec:capecod}
Our fourth example is the CAPE-Cod trial (NCT02517489), which aims to assess the efficacy of hydrocortisone at ICU on patients suffering of severe community-acquired pneumonia.  At the beginning of the COVID-19 pandemic the trial was active and including patients. As SARS-CoV-2 pneumonia was not an exclusion criterion of CAPE-Cod, centers started to include COVID-19 infected patients into the study. The clinical characteristics between the two indications differed, so trial stakeholders have decided to put temporarily on hold the inclusions in CAPE-Cod study and to use the information of COVID-19 patients by embedding a specific study considering COVID-19 indication only. A group-sequential design using the alpha-spending approach by Kim-DeMets \citep{KimDeMets1987a, KimDeMets1987b} was chosen for the COVID-19 substudy to account for the considerable uncertainty with regard to the treatment effect in this new group of patients. If the CAPE-Covid study does not achieve the required sample size or stop (for efficacy or futility) before next autumn, there will potentially be inclusions of patients into two studies, as community-acquired pneumonia is a seasonal disease and COVID-19 will still be present. Taking into account patients' heterogeneity will be a major methodological challenge for this trial.

\section{How adaptive designs might be used to overcome COVID-19 challenges} \label{sec:general}
In this section, guidance is provided on (i) where blinded adaptations can help; (ii) how to achieve type I error rate control, if required, with unblinded adaptations; (iii) how to deal with potential treatment-effect heterogeneity; (iv) how to utilize early read-outs; and (v) how to utilize Bayesian techniques.

\subsection{Where adaptations based on blinded data can help} \label{sec:blinded}
With blinded data we mean here more generally non-comparative data, i.e. data pooled across treatment arms \citep{FDA2019}. Although the trial could be open, the adaption could be informed by blinded data in the sense that they are non-comparative. Generally speaking, potential inflation of type I error rate is less of a concern when adaptations are informed by blinded data \citep{EMA2007, FDA2019}. Therefore, they might be considered first before looking into unblinded adaptations with knowledge of treatment effect estimates.

In the introduction an outline of the potential challenges for clinical trials by the COVID-19 pandemic was provided. In order to assess the extent by which a trial is affected by these, blinded data may be interrogated with regard to the following: baseline patient characteristics; premature study or treatment discontinuations; missing data during follow-up; protocol violations; and nuisance parameters of the outcomes including event rates and variances. The findings may be compared be planning assumptions. Furthermore, time trends can be explored in the blinded data and any changes might be attributed to the COVID-19 if these coincide with the onset of the pandemic (see e.g. \citet{FriedeHenderson2003}).

The findings of such blinded analyses might trigger investigations into resizing the trial. The resizing could be based on blinded or unblinded data; appropriate procedures will be considered in Section \ref{sec:resizing}. However, adaptations are not restricted to sample size reestimation but also include other adaptations such as changes in the statistical model or test statistics to be used. For instance, observed changes in baseline characteristics might be reflected in the statistical model by including additional covariates. Similarly, findings regarding missing data, e.g. due to missed visits, might suggest to adopt a more robust analysis approach.

\subsection{With unblinded adaptations: Methods to control type I error rate} \label{sec:type1error}
It is well known that repeated analyses of accumulating clinical trial data can lead to estimation bias and to inflation of the type I error rate \citep{Armitage1969}.  For this reason there is generally a reluctance to modify the design of a clinical trial during its conduct for fear that the scientific integrity will be compromised.  The necessity of a severe pause in recruitment in many trials due to the current pandemic, however, raises questions of whether additional analyses can be added to an ongoing trial to enable the data obtained so far to be analysed now, with a decision of whether or not to continue with the trial at a later post-COVID-19 time.  Although the current situation of clinical trials being conducted in the setting of a global pandemic is without precedent, the particular question of adding interim analyses to a trial is not a new one.

\citet{ProschanHunsberger1995} introduced the concept of a conditional error function specified prior to the first analysis of accumulating data to be a function that gives the conditional probability of a type I error given the stage 1 data, summarized by a standardized normal test statistic, $z_1$.  In order to control the type I error of the test at level $\alpha$, the conditional error function, $A(z_1)$, with range $[0,1]$, must have
\[
\int_{-\infty}^\infty A(z_1) \phi(z_1) dz_1 = \alpha
\]
where $\phi$ is the standard normal density function. \citet{Wassmer1998} and \citet{MullerSchafer2001} showed how this approach can be used to change a single-stage trial to have a sequential design equivalent to that obtained using a group-sequential or combination function test.

The conditional error principle thus enables a trial planned with a single final analysis to be modified at any point prior to that analysis to have a sequential design, with this constructed in such that the type I error rate is not inflated. It should be noted, however, that it is necessary to specify how any data before and after the interim analysis are combined before the first interim analysis is conducted.  Modification of the design to include initially unplanned interim analyses will also generally lead to a reduction in the power of the trial, as considered in more detail below.

A similar application of the conditional error principle can be used to modify a trial initially planned with interim analyses. For ongoing clinical trials initially planned with interim analyses, the impact of the COVID-19 pandemic may lead to a desire to modify the timing of the planned analyses. Analyses are often taken at times specified in terms of the information available, which may be proportional to the number of patients for a normally distributed endpoint, or to the number of events for a time-to-event endpoint, or given by the number of events for a binary endpoint.  Changes to the timing of the interim analyses do not generally lead to an inflation of the type I error rate provided these are not based on the observed treatment difference, and the spending function method \citep{LanDeMets1983} can be used to modify the critical values used to allow for such changes.  If the timing of interim analyses is based on the estimated treatment difference the type I error rate can be inflated using a group-sequential test (see, for example, \citet{Proschan1992}).  The combination testing approach could be used to control the type I error rate in this setting (see, for example, \citet{Brannath2002}).

\subsection{Treatment-effect heterogeneity} \label{sec:heterogeneity}
Homogeneity over the stages of a multistage design has always been a topic of discussion. Even without pandemic disruptions, there are various reasons why studies could change over time: Some sites may only contribute to part of study, the study population may change over time, e.g. for reasons of a depleted patient pool, and the disease under study itself may vary over time. While many of these reasons also apply to fixed sample size designs, multistage and especially adaptive trials are under obligation to deliver justifications of why the stages can be considered sufficiently homogeneous in order to test a common hypothesis. The EMA reflection paper on adaptive designs states that ``Using an adaptive design implies [...] that methods for the assessment of homogeneity of results from different stages are pre-planned'' \citep{EMA2007}. One option they give is the use of heterogeneity tests as known from the area of meta-analyses. However, as \citet{FriedeHenderson2009} point out, this can reduce the power of studies substantially even in case of no heterogeneity: Such tests are typically carried out at a higher significance level than the standard ones, thus accepting a higher false positive rate. An addition they bring forward is searching for timewise cutpoints in the data. Conclusions about the relationship between timing of change and occurrence of interim analyses can then be drawn from the resulting findings.
In the current COVID-19 situation, the challenge statisticians face is similar to the general challenge described above. The nature and the severity of the impact will very much depend on the actual situation of the trial and the disease under study. Consequently, the way to deal with them may differ as described elsewhere in this paper. Here, we will focus on the question of whether the COVID-19-related changes are such that a rescue by introducing an adaptive design seems justifiable from the homogeneity aspect. There is one major difference to the situation described in the preceding paragraph: The presence of one or two cutpoints, depending on whether the trial will continue both during and after COVID-19, can be taken as a given. Also the question of whether the changes are due to a possibly performed interim analysis or due to COVID-19 seems moot; the question we need to answer is whether a combination is justified.

In some cases, it will be obvious that a combination is not warranted. One example for such a case could be studies in respiratory diseases with hospitalizations included in the endpoint, where a COVID-19 related hospitalization may be an intercurrent event. In other cases, it may not be that obvious and there might be reasons to believe that the pooled patient set is suitable to answer the study hypothesis. Due to the reasons listed above, again a formal heterogeneity test will not be the tool of choice. The EMA Draft Points to Consider on COVID-19 \citep{EMA2020b} does not make mention of the burden of proving homogeneity; rather it states the need of ``additional analyses [...] to investigate the impact of the three phases [...] to understand the treatment effect as estimated in the trial''. While this does not give sponsor carte blanche to combine as they wish, it clearly leaves room for a number of approaches of justifying combination, both from a numerical and a medical perspective. The estimand framework will be an important factor in the decision on pooling or not pooling the data as it will make arguments visible in a structured way: If estimands differ between study parts, then no meaningful estimator for them will be obtained from pooled data (see also Section \ref{sec:regulatory}).

What can statistical methodology contribute if it must be conceded that pooling the patients is not justifiable? In some situations, the number of patients before the COVID-19 impact may already be sufficient to provide reasonable power (see Section \ref{sec:almost_done}). In this case, also patients in the COVID-19 timeframe would need to be analyzed, but it is unclear how they might be included. General guidance for such patients is given as repeating the analysis including all patients and discuss changes in the treatment effect estimate. Medical argumentation will then be needed to underpin the assumptions that changes are due to COVID-19. In some cases, causal inference can help estimate outcomes from those patients under the assumption that COVID-19 had not happened. If interested in the treatment effect in a pandemic free world, it might be worth clarifying the question of interest by relating to the estimand framework \citep{ICHE9R1} where COVID-19 is seen as an intercurrent event. Alternatively, one could standardize results from all patients to the subgroup of patients pre-COVID-19 (e.g., \citet{ShuTan2018} and \citet{HernanRobins2020}). Sometimes, also an artificial censoring at the COVID-19 impaction and the use of short-term information (see Section \ref{sec:early}) to estimate final outcomes will provide a helpful sensitivity analysis.

If it is not feasible to gain sufficient evidence from the pre-COVID-19 patients and a combination does not seem justifiable, then it may be advisable to pause the trial and to re-start it after the COVID-19 time. The during-COVID-19 patients should be included in supporting analyses, but the main evidence will come from the patient pool not directly affected by the pandemic (see also \cite{Anker2020}). Short-term endpoints from during-COVID-19 patients may be used in addition to completed patients to inform decisions on the future sample size.

\subsection{Use of early read-outs} \label{sec:early}
The efficiency of adaptive designs depends on the timing and frequency of interim analyzes. Decision making in adaptive study designs requires availability of information on the endpoints utilized for decision making for a sufficient number of patients. Therefore, care should be taken when conducting an unplanned (or early) interim analysis in trials affected by COVID-19. Focusing on the primary endpoint data only will routinely exclude the many individuals for whom the primary endpoint is not available and might therefore potentially be misleading. In particular, this is relevant to studies interrupted by COVID-19 as investigators may already wish to assess the treatment effect using the pre-pandemic data.
Several proposals have been made to use the information on early read-outs to inform the adaptation decision \citep[e.g.][]{Friede2011, Rufibach2016, Jorgens2019}. Although the information is different from the primary outcome with all limitations that this might have \citep{Zackin1998}, a greater proportion of subjects can contribute to the analysis. This is especially useful in trials where only information about the short-term endpoint would be available at the interim analysis \citep{Friede2011}. 

If primary endpoint data are available, another approach is to retain the pre-specified long-term endpoint as the primary focus of the interim analysis, but to support it with  information on short-term data. In particular, such methodology exploits the possible statistical association between the short- and long-term endpoints to provide information about the long-term primary endpoint on patients who didn't reach their primary endpoint yet. 
This provides an efficient compromise between the less efficient approach of using only information on subjects who have been followed through to the long-term primary endpoint, and the potentially misleading approach of using only short-term information.

A likelihood-based estimator for binary outcomes that combines short- and long-term data assessed at two timepoints was discussed by \cite{MarschnerBecker2001}. 
They use a likelihood function that depends on three binomial distributions, which model the probability of success at both timepoints as well as the conditional probability of success at the last time point given the short-term endpoint. The maximum likelihood estimator for the probability of a successful primary endpoint in each treatment arm is
\[
\hat{P}(L_j=1\mid S_j=1)\hat{P}(S_j=1)+\hat{P}(L_j=1\mid S_j=0)(1-\hat{P}(S_j=1)),
\]
where $L_j= \{0, 1\}$ and $S_j= \{0, 1\}$ denote the outcome in the trial observed on respectively the long-term primary endpoint and the short-term endpoint in treatment arm $j\in\{0,1\}$.
Implementing their estimator for the treatment difference in the conditional power approach improves decision making at an interim analysis both for futility stopping and sample size re-assessment \citep{Niewczas2019}. To preserve the type I error the combination test with final test statistic (see Section \ref{sec:type1error}),
\[
\sqrt{w}z_1+\sqrt{1-w}z_2,
\]
can be used.
Here, the first stage test statistic $z_1$ is defined by the cohort of patients included before the interim analysis, but the data used comes from the primary endpoint.
For patients who were included before the COVID-19 impact but for whom the primary endpoint was observed after that impact, this would mean that their primary endpoints would need to be analysed together with those occurring before the impact. A disadvantage of this test procedure is that it prohibits early stopping for efficacy.
In situations where it is not feasible to obtain sufficient information from the pre-COVID-19 patients, it may be advisable to support the interim analysis with historical data \citep{VanLancker2019}. \cite{Sooriyarachchi2006} proposed a score test for incorporating binary assessments taken at three fixed time points. 
\cite{GalbraithMarschner2003} expanded the likelihood-based approach described in \cite{MarschnerBecker2001} to continuous endpoints assessed at an arbitrary number of follow-up times with a proposition of extending the topic to group sequential designs or conditional power approaches. This approach was generalized by \cite{HampsonJennison2013} to a group sequential design with delayed responses.

A shortcoming of these approaches is that they do not take into account baseline covariates in order to make fully efficient use of the information in the data. In addition, the incorporation of baseline covariates and longitudinal measurement of the clinical outcome is likely to involve a heavier dependence on statistical modelling, which raises concerns that their misspecification may result in bias. This problem is overcome by \cite{VanLancker2020}. They propose an interim procedure that is applicable for binary and continuous outcomes assessed at an arbitrary number of follow-up times and that allows the incorporation of baseline covariates in order to increase efficiency. They realise this by considering the estimation of the treatment effect at the time of the interim analysis as a missing data problem. This problem is addressed by employing specific prediction models for the long-term endpoint which allow, besides multiple short-term endpoints, the incorporation of baseline covariate information. 
When measurements are taken at three different timepoints (baseline, intermediate and final), the probability of a successful primary endpoint in each treatment arm is estimated by
\begin{enumerate}
	\item fitting a model for $L_j$ on $S_j$ and the baseline covariates in the cohort of patients for whom all data is available, and using the model to make predictions $\hat{L}_j$ for all patients for whom $S_j$ is observed
	\item fitting a model for $\hat{L}_j$ on the baseline covariates in the cohort of people  for whom $S_j$ is observed,
	\item taking the mean of the predictions based on the model in Step $2$ for all patients recruited at the time of the interim analysis.
\end{enumerate}
Implementing their estimator for the treatment difference in the conditional power approach improves decision making at an interim analysis both for futility stopping and sample size re-assessment without compromising the type I error rate, even if the prediction models are misspecified. 

The method differs from the previous ones as the proposed interim test statistic, denoted by $z_\tau$, is used as the first-stage test statistic $z_1$. This is allowed as they provide a second stage test statistic that is independent of their proposed interim test statistic $z_\tau$,
\[
z_2=\frac{z-\sqrt{\tau}z_\tau}{\sqrt{1-\tau}},
\]
where $z$ is the final test statistic based on primary endpoint data only and $\tau$ is the information fraction at the time of the interim analysis.
This allows for statistical hypothesis testing at the interim analysis.
However, the type I error will be compromised if information other than the current test statistic is used for interim decisions. In that case, the type I error can be preserved by defining the first stage $p$-value by the cohort of patients included before the interim analysis \citep[e.g.][]{Friede2011, Jenkins2011}. 

Similarly, care has to be taken when applying flexible study designs to time to event data \citep{Bruckner2018}. When data are separated into stages by the occurrence of the primary event, the type I error will be compromised if information other than the current logrank test statistic is used for interim decisions \citep{Bauer2004}. If short-term endpoints are to be used, \cite{Jenkins2011} proposed to base the separation on patients instead of on events. Similar as for other endpoints, this would mean that the primary event for patients who were included before the COVID-19 impact but did experience their primary event only after that impact, would need to be analysed together with those occurring before the impact. Depending on the actual impact, it may be appropriate to either use these patients as a separate cohort -- in which case their short-term endpoint should not be used for decision making -- or to artificially censor them at the impact timepoint and use their complete data for supplemental analyses only.

In addition, the described methodology is also appropriate for interim analyses that take into account post-COVID-19 data if one can assume that the missing mechanism due to COVID-19-related drop-out is missing completely at random. In the Appendix of \cite{VanLancker2020} an extension of their method that allows the weaker assumption that missingness is at random is discussed.  An alternative for the other methods is to consider more detailed informative missingness models (e.g., via multiple imputation \citep{Sterne2009}).

\subsection{A Bayesian perspective}
Inter-patient heterogeneity as well as intra-patients heterogeneity are both very common in clinical trials. In COVID-19, however, several types of heterogeneity might add to the ususal level of variability. These include (1) patients infected or not by COVID-19, especially incidence of COVID-19 variability per country in international multi-center trials, (2) patients’ outcomes (in cancer studies is the present mortality due to the disease or immunosuppressed systems), (3) patients’ follow-up, and (4) patients’ compliance due to missing treatments. One way of considering these types of heterogeneity is to use Bayesian approaches during, if possible at all, or at the end of the trial. Using hierarchical Bayesian methods associated with Bayesian evidence synthesis methods will allow to account for different type of heterogeneity \citep{Friede2017, Rover2016, Thall2008}. These approaches take into account uncertainty in estimating the between-trial or subgroup heterogeneity but they can also be used in the setting of within-trial heterogeneity. By using potential variation of the scale parameter of the heterogeneity prior would facilitate sensitivity analyses. \citet{Friede2017} proposed a Bayesian random-effects meta-analyses with priors covering plausible heterogeneity values. In the setting of within-trial heterogeneity prior calibration of each source of heterogeneity is at most interest, indeed one should not be limited to methods accounting for only one source of heterogeneity as more than one type can be present. Let $\phi$ be the within-trial standard deviation, it determines the degree of heterogeneity across patients either included before or after COVID-19 pandemic or patients infected or not by COVID-19 (or any other COVID-19 source of heterogeneity) and $\mu$ the parameter of interest. Under Bayesian inference, uncertainty for $\phi$ is automatically accounted for and inference for $\mu$ and $\phi$ can be captured by the joint posterior distribution of the two parameters. The key point is in the choice of the prior distribution of $\phi$, in particular when subgroups are small or unbalanced. In the absence of relevant external data or information about within-trial heterogeneity, the 95\% prior interval of $\phi$ should capture small to large heterogeneity. Moreover, the use of a Bayesian approach entails the question of what constitutes sensible prior information in the context of COVID-19 in which there is a continual updating of information that is still not considered reliable. This may be argued on the basis of the endpoint in question, that is, what is the plausible amount of heterogeneity expected, what constitutes relevant external data, and how this information may be utilized. A relatively simple solution would be the use of weakly informative priors.  For priors of effect parameters, adaptive priors using power or commensurate prior approaches have proved to be efficient in updating if, when and how to incorporate external information \citep{Hobbs2012, Ollier2019}.

\section{Resizing the trial} \label{sec:resizing}
In the following we develop approaches to resizing a clinical trial affected by the COVID-19 pandemic. Specifically, we consider to stop a trial early, the use of group-sequential designs, and sample size adjustment. All methods are implemented in a freely available R shiny app, which is briefly introduced in Section \ref{sec:app}.

\subsection{Almost done: To stop or not to stop early}\label{sec:almost_done}

In case data collection is nearly finished at the time of the COVID-19 impact, a natural question that comes into mind is whether one should analyze the trial early based on the data collected so far accepting some loss in power. In the following, we focus on superiority trials comparing a treatment versus placebo (or standard treatment) with allocation ratio $1:r$ for placebo versus treatment. Let $\alpha$ denote the one-sided significance level and $1-\beta$ the desired power at the planning stage of the trial.

Assume that the endpoint of interest follows a normal distribution. Let $\delta$ denote the assumed difference between the means under the alternative hypothesis and let $\sigma^2$ denote the common variance for both arms. The total sample size $N$ needed to achieve a desired power of $1-\beta$ is then given by
\begin{align}
N =& \left(z_{1-\alpha} + z_{1-\beta}\right)^2 \frac{\sigma^2}{\delta^2} \frac{(r+1)^2}{r}
\label{eqn:samplesize}
\end{align}
with $z_{1-\alpha}$ and $z_{1-\beta}$ denoting the $1-\alpha$- and $1-\beta$-quantile of the cumulative standard normal distribution. We assume that the trial was originally planned to be analyzed based on $N$ observed patients but so far has only data available for $n=\tau N$ patients. Solving Equation (\ref{eqn:samplesize}) for $1-\beta$ and replacing $N$ with $n$ yields the power if the trial is analyzed early based on the data observed:
\begin{align}
1-\beta_n =& \Phi\left(\frac{\delta}{\sigma} \sqrt{\frac{r}{(r+1)^2}} \sqrt{n} - z_{1-\alpha}\right)
\end{align}
where $\Phi(\cdot)$ denotes the cumulative distribution function of the standard normal distribution. However, as $n=\tau N = ( \tau (z_{1-\alpha} + z_{1-\beta})^2 \sigma^2 (r+1)^2)/(\delta^2r)$, we can show that the resulting power can be rewritten as
\begin{align}
1-\beta_n =& \Phi\left(\frac{\delta}{\sigma} \sqrt{\frac{r}{(r+1)^2}} \sqrt{n} - z_{1-\alpha}\right)
= \Phi\left(z_{1-\beta}\sqrt{\tau} - z_{1-\alpha} (1-\sqrt{\tau})\right)
\label{eqn:power_fixed_n}
\end{align}
which is independent of the allocation ratio $r$, the difference between the means $\delta$ and the variance $\sigma^2$. Instead the power based on $n$ available patients only depends on the fraction $\tau$ as well as the significance level $\alpha$ and the desired power $1-\beta$ at planning stage.

\begin{figure}
\centering
\includegraphics[width=0.99\textwidth, angle = 0]{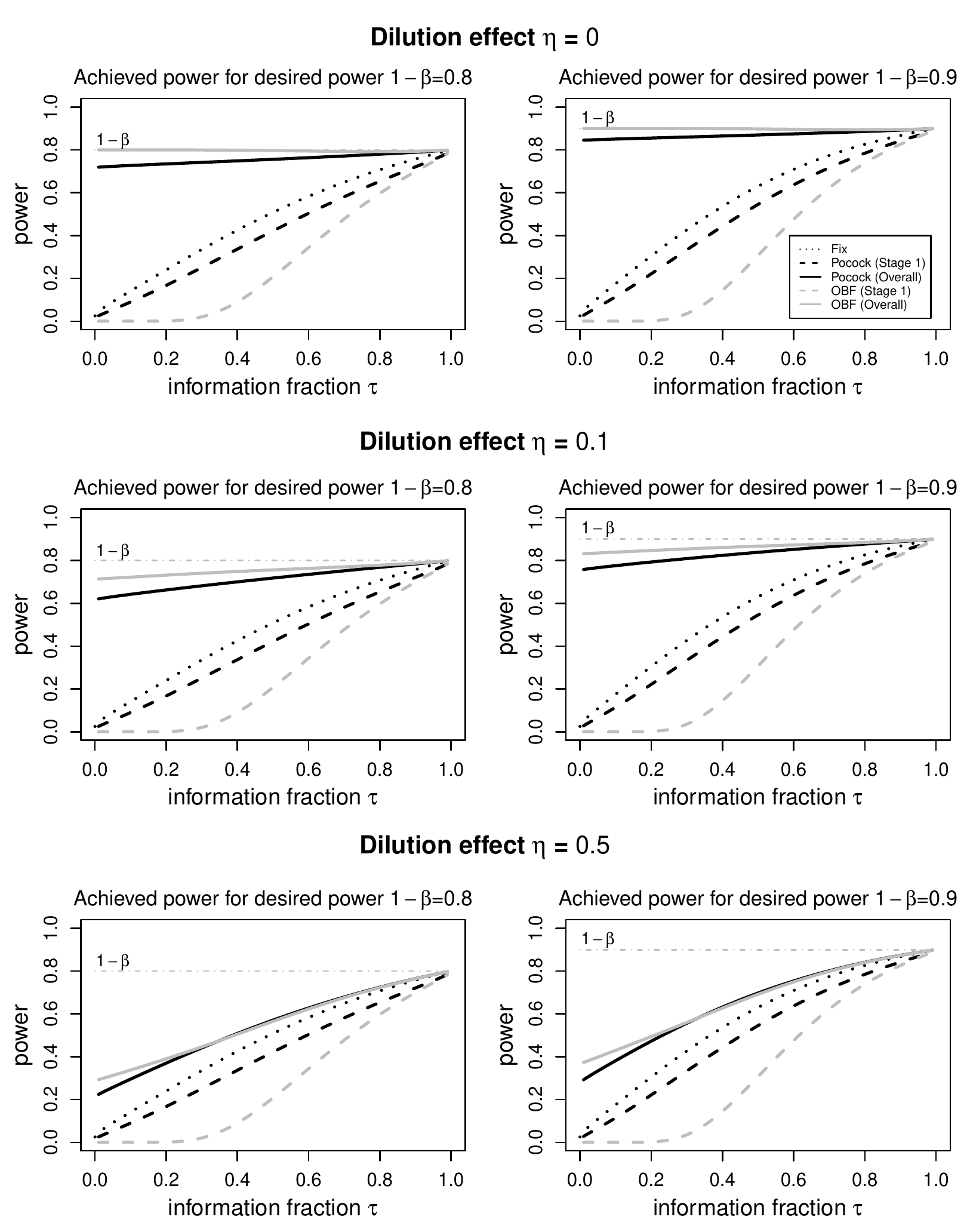}
\caption{Resulting power depending on the information fraction $\tau$ for a dilution effect of $\eta=0$, $\eta=0.10$, or $\eta=0.50$ for the fixed design (black dotted line), the Pocock group sequential design (stage 1: black dashed line, overall: black solid line), and the O'Brien-Fleming group sequential design (stage 1: gray dashed line, overall: gray solid line) for a desired power of either $1-\beta=0.80$ or $1-\beta=0.90$.}
\label{fig:dilution_1}
\end{figure}

Resulting values for the power depending on the information fraction $\tau$ are shown in Figure \ref{fig:dilution_1} (black dotted line) and Table \ref{tab:tau_power}. For a desired power of $1-\beta=0.80$, if data is available for about 80\% ($\tau=0.80$) of the planned patients, the absolute loss in power for the fixed design is about 10 percentage points while for a planned power of $1-\beta=0.90$, the loss in power is about 7 percentage points.

\begin{table}[!ht]
\caption{Resulting values for the power based on $n$ patients ($1-\beta_n$) depending on the fraction $\tau$ of data already available for an originally planned power of $1-\beta=80\%$ or $1-\beta=0.90$ for a dilution effect of $\eta=0$ or $\eta=0.10$.} \label{tab:tau_power}
\begin{center}
\begin{tabular}{r|r|rr|rr|r|rr|rr}
\hline
&& \multicolumn{4}{c|}{$\boldsymbol{1-\beta=0.80}$}  &&  \multicolumn{4}{c}{$\boldsymbol{1-\beta=0.90}$} \\
&& \multicolumn{2}{c}{Pocock} &  \multicolumn{2}{c|}{OBF*} && \multicolumn{2}{c}{Pocock} &  \multicolumn{2}{c}{OBF*} \\
&& \multicolumn{2}{c}{Stage} &  \multicolumn{2}{c|}{Stage} && \multicolumn{2}{c}{Stage} &  \multicolumn{2}{c}{Stage} \\
$\boldsymbol{\tau}$ &  fix & 1  & 2 & 1  & 2 &  fix & 1  & 2 & 1  & 2 \\
\hline
\multicolumn{11}{c}{$\boldsymbol{\eta = 0, \psi = 1}$}  \\
\hline
0.50 & 0.508 & 0.422 & 0.756 & 0.207 & 0.797 &  0.630 & 0.545 & 0.870 & 0.307 & 0.898\\
0.60 & 0.583 & 0.504 & 0.764 & 0.344 & 0.795 &  0.709 & 0.637 & 0.875 & 0.476 & 0.896\\
0.70 & 0.650 & 0.581 & 0.772 & 0.478 & 0.793 &  0.774 & 0.717 & 0.880 & 0.622 & 0.895\\
0.80 & 0.707 & 0.653 & 0.780 & 0.597 & 0.792 &  0.826 & 0.785 & 0.886 & 0.739 & 0.895\\
0.85 & 0.733 & 0.688 & 0.785 & 0.650 & 0.793 &  0.848 & 0.815 & 0.889 & 0.786 & 0.895\\
0.90 & 0.757 & 0.721 & 0.789 & 0.699 & 0.794 &  0.868 & 0.842 & 0.892 & 0.826 & 0.896\\
0.95 & 0.780 & 0.754 & 0.794 & 0.745 & 0.796 &  0.885 & 0.868 & 0.896 & 0.862 & 0.897\\
0.99 & 0.796 & 0.785 & 0.799 & 0.783 & 0.799 &  0.897 & 0.890 & 0.899 & 0.889 & 0.899\\
\hline
\multicolumn{11}{c}{$\boldsymbol{\eta = 0.1, \psi = 1}$}  \\
\hline
0.50 & 0.508 & 0.422 & 0.718 & 0.207 & 0.756 &  0.630 & 0.545 & 0.838 & 0.307 & 0.867\\
0.60 & 0.583 & 0.504 & 0.735 & 0.344 & 0.763 &  0.709 & 0.637 & 0.852 & 0.476 & 0.872\\
0.70 & 0.650 & 0.581 & 0.752 & 0.478 & 0.770 &  0.774 & 0.717 & 0.864 & 0.622 & 0.878\\
0.80 & 0.707 & 0.653 & 0.768 & 0.597 & 0.778 &  0.826 & 0.785 & 0.877 & 0.739 & 0.884\\
0.85 & 0.733 & 0.688 & 0.776 & 0.650 & 0.783 &  0.848 & 0.815 & 0.883 & 0.786 & 0.887\\
0.90 & 0.757 & 0.721 & 0.784 & 0.699 & 0.788 &  0.868 & 0.842 & 0.888 & 0.826 & 0.891\\
0.95 & 0.780 & 0.754 & 0.792 & 0.745 & 0.793 &  0.885 & 0.868 & 0.894 & 0.862 & 0.895\\
0.99 & 0.796 & 0.785 & 0.798 & 0.783 & 0.798 &  0.897 & 0.890 & 0.899 & 0.889 & 0.899\\
\hline
\multicolumn{11}{l}{* OBF: O'Brien-Fleming}
\end{tabular}
\end{center}
\end{table}

\subsection{Blinded sample size reestimation}
In Section \ref{sec:blinded} we touched upon blinded adaptations. Here we make some comments specifically on blinded sample size reestimation. Blinded sample size reestimation procedures are well established to account for misspecifications of nuiscance paramters in the planning phase of a trial \citep{FriedeKieser2013}. In this situation considered, namely the impact of the COVID-19 pandemic, a number of circumstances might make a resizing of the trial necessary and, as discussed, could be addressed in a blinded sample size review. In particular, censoring of follow-up might make it necessary to assess the sample size and lenght of follow-up. This would be of particular importance in long running trials, particularly prevalent in  chronic conditions. In the context of heart failure trials, \citet{Anker2020} suggested to censor observations due to regional COVID-19 outbreaks. Such actions would imply a resizing of the trial, potentially in terms of number of patients recruited and length of follow-up, to maintain previously set or in the light of the pandemic revised timelines \citep{Friede2019}.

\subsection{Switching from a fixed to a group-sequential design}\label{sec:gsd}
Assume that a fixed design was planned with $N$ patients to be analyzed. However, only $n$ patients have been observed so far. The question is now, whether we can change to a group-sequential design (GSD) using the $n$ patients for the first stage while the final sample size is still $N$. That is, the trial is analyzed as a group sequential design using the total sample size from the fixed design. The difference between Section \ref{sec:almost_done} and the situation here is, that we will adjust the critical value to allow for two tests of the null hypothesis.

Let $c_1$ and $c_2$ denote the critical values for a two-stage design and let $\tau=n/N$ denote fraction of data being used for the first stage. The variance-covariance matrix for the two test statistics for the first stage and the final analysis is given by
\begin{align}
\sigma^2 =
\begin{pmatrix}
1 & \sqrt{\tau} \\
\sqrt{\tau} & 1
\end{pmatrix}.
\end{align}

Using $\Phi$ to denote the cumulative distribution function of the bivariate standard normal distribution the type I error rate is then given by
\begin{align}
\alpha = 1- \Phi\left(
\begin{pmatrix} c_1 \\ c_2\end{pmatrix},
\mu=\begin{pmatrix} 0\\ 0\end{pmatrix},
\sigma^2 = \begin{pmatrix}
1 & \sqrt{\tau} \\
\sqrt{\tau} & 1
\end{pmatrix}
\right).
\end{align}
and the power is given by
\begin{align}
1-\beta = 1- \Phi\left(
\begin{pmatrix} c_1 \\ c_2\end{pmatrix},
\mu=\begin{pmatrix} \sqrt{\tau} \frac{\delta}{\sigma}\sqrt{\frac{r}{(r+1)^2}}\sqrt{N} \\ \frac{\delta}{\sigma}\sqrt{\frac{r}{(r+1)^2}}\sqrt{N} \end{pmatrix},
\sigma^2 = \begin{pmatrix}
1 & \sqrt{\tau} \\
\sqrt{\tau} & 1
\end{pmatrix}
\right).
\label{eqn:gsd_power}
\end{align}
Replacing $N$ in Equation (\ref{eqn:gsd_power}) by the right-hand side of Equation (\ref{eqn:samplesize}) yields
\begin{align}
1-\beta
=& 1- \Phi\left(
\begin{pmatrix} c_1\\ c_2\end{pmatrix},
\mu=\begin{pmatrix}
\sqrt{\tau} \frac{\delta}{\sigma}\sqrt{\frac{r}{(r+1)^2}}\sqrt{\left(z_{1-\alpha} + z_{1-\beta}\right)^2 \frac{\sigma^2}{\delta^2} \frac{(r+1)^2}{r}} \\
\frac{\delta}{\sigma}\sqrt{\frac{r}{(r+1)^2}}\sqrt{\left(z_{1-\alpha} + z_{1-\beta}\right)^2 \frac{\sigma^2}{\delta^2} \frac{(r+1)^2}{r}} \end{pmatrix},
\sigma^2 = \begin{pmatrix}1 & \sqrt{\tau} \\ \sqrt{\tau} & 1 \end{pmatrix}
\right)
\nonumber\\
=& 1- \Phi\left(
\begin{pmatrix} c_1 \\ c_2\end{pmatrix},
\mu=\begin{pmatrix} \sqrt{\tau} \left(z_{1-\alpha} + z_{1-\beta}\right) \\ \left(z_{1-\alpha} + z_{1-\beta}\right) \end{pmatrix},
\sigma^2 = \begin{pmatrix}1 & \sqrt{\tau} \\ \sqrt{\tau} & 1 \end{pmatrix}
\right)
\label{eqn:power_gsd}
\end{align}

As in Section \ref{sec:almost_done}, the resulting power does only depend on the values for the significance level $\alpha$, the desired power $1-\beta$ at planning stage and the fraction of data available at interim $\tau$. It also depends on the critical values chosen to control the type I error rate.

The top row of Figure \ref{fig:dilution_1} shows the resulting power depending on the information fraction $\tau$ for a planned desired power of either $1-\beta=0.80$ (left-hand panel) or $1-\beta=0.90$ (right-hand panel). 

Table \ref{tab:tau_power} lists the resulting power for some values of $\tau$ for a desired power of either 80\% oder 90\%. The first column gives the value for $\tau$, columns 2 to 6 give the resulting power for the fixed design as well as for both stages the Pocock and the O'Brien-Fleming design for a desired power of $1-\beta=0.80$, and columns 7 to 11 give the achieved power for a desired power of $1-\beta=0.90$. The first set of lines assume that there is no dilution effect (see Section \ref{sec:dilution}) for patients enrolled into the trial after the COVID-19 outbreak while the second set assumed that the dilution effect is $\eta=0.10$. For example, if 80\% of the planned data has been collected before the COVID-19 outbreak, the resulting power for a fixed design is 0.707 if the planned power is $1-\beta=0.80$. Using a Pocock GSD, the power for the first stage is 0.653 while the overall power at the end of the second stage is 0.78. For the O'Brien-Fleming GSD, the power for the first stage is 0.597, while the overall power is 0.792.

\subsection{Dilution effect $\eta$} \label{sec:dilution}
Another likely scenario is that due to the COVID-19 outbreak the response to the treatment has changed and maybe even the response to the control treatment. Let $\mu_{c0}$ and $\sigma_{c0}^2$ denote the mean and the variance for the control group before the outbreak and let $\mu_{c1}$ and $\sigma_{c1}^2$ denote the mean and variance for the control group after the outbreak. Analogously, the means and variances for the treatment group before and after the outbreak are denoted with $\mu_{t0}$, $\mu_{t1}$, $\sigma_{t0}^2$, and $\sigma_{t1}^2$. Let $\delta=\mu_{t0}-\mu_{c0}$ denote the treatment effect before the outbreak started. The difference between the means after the outbreak started can then be expressed as a fraction of the difference before the outbreak started, i.e. $\mu_{t1}-\mu_{c1}=(1-\eta)\delta$. In the following, $\eta$ will be called the dilution effect. The patients are randomized to control and treatment in a $1:r$ ratio. Assume that at the time of the outbreak $n=\tau N$ patients have been enrolled into the trial and that it is planned to enroll a total of $N$ patients. Let $t_0$ denote the test statistic based on only the patients enrolled before the outbreak and let $t_1$ denote the test statistic based on only the patients enrolled after the outbreak. Furthermore, let $t$ denote the test based on all enrolled patients.

With respect to the change of the means after the outbreak, two possible definitions can be thought of
\begin{enumerate}
\item relative change of means: $\mu_{c1} = (1-\eta_c) \mu_{c0}$ and $\mu_{t1} = (1-\eta_t) \mu_{t0}$,
\item absolute change of means: $\mu_{c1} = \mu_{c0}-\epsilon_c$ and $\mu_{t1} = \mu_{t0}-\epsilon_t$.
\end{enumerate}
Using $\epsilon_c=\eta_c\mu_{c0}$ and $\epsilon_t=\eta_t\mu_{t0}$ (or alternatively, $\eta_c=\epsilon_c/\mu_{c0}$ and $\eta_t=\epsilon_t/\mu_{t0}$), it can be shown that both approaches can be converted into one another. For the variances, we only consider a relative change of the variance and define $\sigma_{c1}^2 = \psi_c \sigma_{c0}^2$ and $\sigma_{t1}^2 = \psi_t \sigma_{t0}^2$.

A common assumption is that the variances for the treatment and the control group are the same. Here, we consider the case of $\sigma_{t0}^2=\sigma_{c0}^2=\sigma_0^2=\sigma^2$ and $\sigma_{t1}^2=\sigma_{c1}^2=\sigma_1^2$ with $\psi_t=\psi_c=\psi$ and $\sigma_1^2 = \psi \sigma_0^2$. That is, we assume equal variances for the two arms but not necessarily equal variances before and after the outbreak. The joint distribution of $t_1$, $t_2$, and $t$ is then given by
\begin{align}
\begin{pmatrix}t_0 \\ t_1 \\ t \end{pmatrix} \sim N &\left(
\begin{pmatrix}
\sqrt{\frac{Nr\tau}{(r+1)^2}} \cdot \frac{\delta}{\sigma} \\
\sqrt{\frac{Nr(1-\tau)}{(r+1)^2}} \cdot \frac{\left(1-\eta\right)}{\sqrt{\psi}}\cdot \frac{\delta}{\sigma} \\
\sqrt{\frac{Nr}{(r+1)^2}} \cdot \frac{\tau+(1-\tau)\left(1-\eta\right)}{\sqrt{\tau + (1-\tau)\psi}} \cdot \frac{\delta}{\sigma}
\end{pmatrix},
 \begin{pmatrix}
1 \\
0 & 1 \\
\sqrt{\frac{\tau}{\tau+(1-\tau) \psi}}
& \sqrt{\frac{(1-\tau)\psi}{\tau+(1-\tau) \psi}} & 1
\end{pmatrix}
\right).
\label{eqn:joint_simple_2}
\end{align}
The general solution for the joint distribution can be found in Appendix \ref{app:joint}.

As before, we assume that the original sample size was planned using a one-sided significance level of $\alpha$ to achieve a desired power of $1-\beta$ based on Equation (\ref{eqn:samplesize}). Replacing $N$ with $ ((z_{1-\alpha} + z_{1-\beta})^2 \sigma^2 (r+1)^2)/(\delta^2r)$ yields
\begin{align}
\begin{pmatrix}t_0 \\ t_1 \\ t \end{pmatrix} \sim N &\left(
\begin{pmatrix}
\left(z_{1-\alpha} + z_{1-\beta}\right) \sqrt{ \tau }\\
\left(z_{1-\alpha} + z_{1-\beta}\right)   \frac{\sqrt{1-\tau}\left(1-\eta\right)}{\sqrt{\psi}} \\
\left(z_{1-\alpha} + z_{1-\beta}\right) \cdot \frac{\tau+(1-\tau)\left(1-\eta\right)}{\sqrt{\tau + (1-\tau)\psi}}
\end{pmatrix},
 \begin{pmatrix}
1 \\
0 & 1 \\
\sqrt{\frac{\tau}{\tau+(1-\tau) \psi}}
& \sqrt{\frac{(1-\tau)\psi}{\tau+(1-\tau) \psi}} & 1
\end{pmatrix}
\right).
\label{eqn:joint_simple_3}
\end{align}

By setting $\psi=1$ (assuming equal variances before and after the outbreak), the equation reduces further to
\begin{align}
\begin{pmatrix}t_0 \\ t_1 \\ t \end{pmatrix} \sim N &\left(
\begin{pmatrix}
\left(z_{1-\alpha} + z_{1-\beta}\right) \sqrt{\tau} \\
\left(z_{1-\alpha} + z_{1-\beta}\right) \sqrt{(1-\tau)} \left(1-\eta\right) \\
\left(z_{1-\alpha} + z_{1-\beta}\right) \left(\tau + (1-\tau)\left(1-\eta\right)\right)
\end{pmatrix},
 \begin{pmatrix}
1 \\
0 & 1 \\
\sqrt{\tau} & \sqrt{1-\tau} & 1
\end{pmatrix}
\right).
\label{eqn:joint_final}
\end{align}
As shown in Sections \ref{sec:almost_done} and \ref{sec:gsd}, the resulting distribution depends on the values for the significance level $\alpha$, the desired power $1-\beta$, and the fraction of data available for the outbreak $\tau$. In the case considered here, the only additional variable is the dilution effect $\eta$.

Figure \ref{fig:dilution_1} shows the resulting values for the power depending on the information fraction $\tau$ for a dilution effect of $\eta=0$, $\eta=0.10$, and $\eta=0.20$. The upper two plots show the achieved power for a dilution effect of $\eta=0$ (see Section \ref{sec:dilution}), the middle plots show the power for a dilution effect of $\eta=0.1$, and the bottom plots for a dilution effect of $\eta=0.5$. The black dotted line gives the resulting values for the power for the fixed design if analyzed early, the black lines give the resulting power for the Pocock design for the first stage (dashed line) and overall (solid line), and the gray lines give the resulting power for the O'Brien-Fleming design for the first stage (dashed line) and overall (solid line). It should be noted that the power for the fixed design as well as the power for the first stage for the GSDs does not change as analyses only uses first stage data which was collected before the outbreak.

In conclusion, if at least 85\% of the data are available and no considerable dilution effect is expected, then the recommendation would be to stop the trial immediately. In all other scenarios, consequences of any decision would need explored carefully using the approaches developed. As we will see in Section \ref{sec:app} below, these are implemented in a R Shiny app to support this process.

\subsection{Sample size adjustment} \label{sec:adjustment}
As shown in Section \ref{sec:dilution}, some loss in power is to be expected if the means and variances for the treatment and control arm change due to the outbreak. In order to regain the desired power of $1-\beta$, the sample size would need to be adjusted. For a fixed design that is analyzed only once, i.e. after all data has been collected from all patients enrolled before and after the outbreak, the sample size for patients enrolled after the outbreak can be calculated as shown below.  Let $n_0$ denote the number of patients already enrolled into the trial before the outbreak and let $\tilde{n}_1$ denote the number to be enrolled after the outbreak started. We wish to determine $\tilde{n}_1$ so that the power based on a total of $\tilde{N} = n_0 + \tilde{n}_1$ enrolled patients is $1-\beta$.

Based on Equation (\ref{eqn:joint_simple_2}), we know that the final test statistic $t$ follows a normal distribution with
\begin{align}
t \sim N &\left(\sqrt{\frac{(n_0+\tilde{n}_1)r}{(r+1)^2}} \cdot \frac{\frac{n_0}{n_0+\tilde{n}_1}+(1-\frac{n_0}{n_0+\tilde{n}_1})\left(1-\eta\right)}{\sqrt{\frac{n_0}{n_0+\tilde{n}_1}+ (1-\frac{n_0}{n_0+\tilde{n}_1})\psi}} \cdot \frac{\delta}{\sigma}, 1\right)
\label{eqn:t_fix}
\end{align}
Solving
\begin{align}
\sqrt{\frac{(n_0+\tilde{n}_1)r}{(r+1)^2}} \cdot \frac{\frac{n_0}{n_0+\tilde{n}_1}+(1-\frac{n_0}{n_0+\tilde{n}_1})\left(1-\eta\right)}{\sqrt{\frac{n_0}{n_0+\tilde{n}_1}+ (1-\frac{n_0}{n_0+\tilde{n}_1})\psi}} \cdot \frac{\delta}{\sigma} - z_{1-\alpha} = z_{1-\beta}
\label{eqn:sample_size_adj}
\end{align}
for $\tilde{n}_1$ yields
\begin{align}
\tilde{n}_1 =& N\tau \frac{\psi - 2 + 2 \tau \eta + \sqrt{\psi^2 - 4\tau(1-\eta)(\eta+\psi-1)}}{\psi - 2 \tau \eta (1-\eta) - \sqrt{\psi^2 - 4\tau(1-\eta)(\eta+\psi-1)}}
\label{eqn:solution_n1_tilde_final}
\end{align}
The derivations can be found in Appendix \ref{app:samplesize_adjustment}. In order to find the sample size for the second part of the trial for a GSD, a search algorithm based on Equation (\ref{eqn:joint_simple_2}) has to be used.

Please note that the dilution effect $\eta$ cannot be estimated from the data, but need to be hypothesized. Of course sensitivity analyzes can be conducted based on different assumptions. The R shiny app introduced in the next section was devised to support such processes.

\subsection{Implementation of the resizing approaches in a R shiny app} \label{sec:app}
\begin{figure}
\centering
\includegraphics[width=0.99\textwidth, angle = 0]{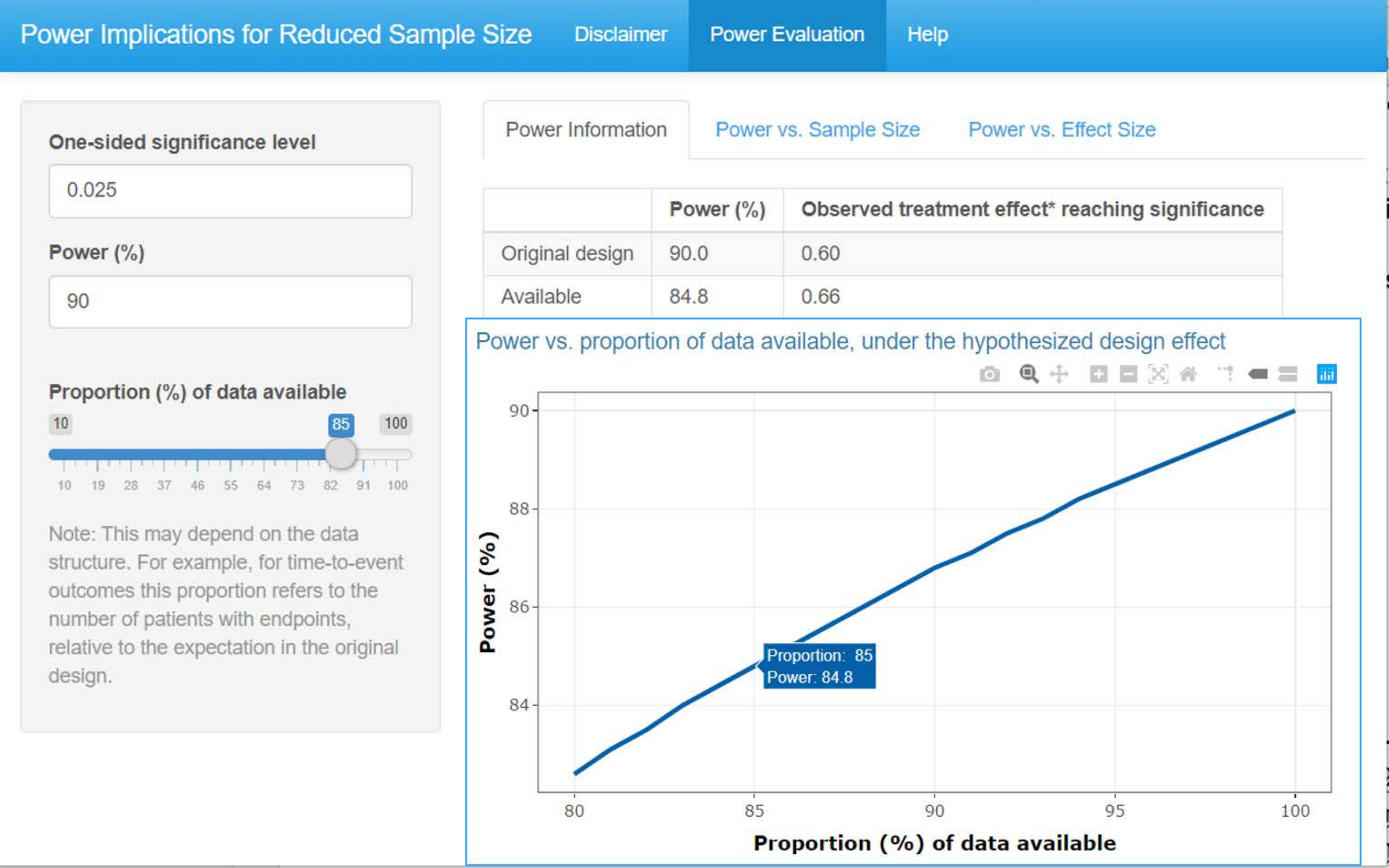}
\caption{Screenshot of the R shiny app.}
\label{fig:app}
\end{figure}

To facilitate the implementation of the proposed methods, an R shiny app was developed as a simple-to-use web-based application. It provides insights into the power properties on the fly, given user-defined input of design parameters. Specifically, it has a module for the calculations shown in Section \ref{sec:almost_done} to answer the following question: If a trial was designed for 90\% power for an assumed treatment effect at a significance level $\alpha=0.025$, what is the power if we conduct the analysis with only 85\% of the patient data? By following \eqref{eqn:power_fixed_n}, the app provides the power (84.8\%) and a plot for different proportions of data available, in addition to 85\%. A screenshot of the app is provided in Figure~\ref{fig:app}.

The app was originally designed to facilitate the discussion by \citet{Akacha2020}, where the same calculation as \eqref{eqn:power_fixed_n} was independently developed. The app is expanded to implement the group sequential design of an interim analysis conducted with data available and a final analysis when the planned data is obtained (see Section \ref{sec:gsd}). Two popular group sequential designs are considered which are the Pocock and the O'Brien-Fleming schemes. In addition, the incorporation of dilution effects allows for more general considerations, as demonstrated in Section \ref{sec:dilution}. Similar outputs as in displayed Figure~\ref{fig:app} are provided with the app for the various scenarios considered above. The app can be accessed at \url{https://power-implications.shinyapps.io/prod/} and comes with a help tab that contains more information about its usage.

\section{Regulatory and operational aspects} \label{sec:regulatory}
The COVID-19 pandemic affects all clinical trials, with implications for studies intended for drug regulation well beyond statistical aspects \citep{EMA2020a, EMA2020b, FDA2020}. For example, on-site monitoring of most trials is suspended during the lockdown and with the interdiction of non-essential travels the recording of adverse events might not be as good if a site visit is replaced by a telephone consultation, or a local laboratory was used instead of the central laboratory. Similarly, the mode of administration of a patient reported outcomes questionnaire might have been changed from an electronic collection at the site on a tablet computer to a paper based version mailed to the patient's home. All these examples may lead to a reduced quality of the trial data which may need to be taken into consideration when interpreting the trial.

As much remains to be learned on the COVID-19 disease manifestations, treatments and pandemic distribution, it appears necessary to monitor the status and integrity of the trial on an ongoing basis. However, it may not be clear in some situations how this can be done in a way that protects the integrity of trial conduct. Care has to be taken that the original responsibilities of a DMC are not expanded beyond reasonable limits. Many of the responsibilities arising during the pandemic might more naturally seem to belong to trial management personnel, as the associated issues can often be addressed adequately without access to unblinded data; this might involve sponsor personnel, steering committees, etc. If important decisions are advised by unblinded results, then of course this should be done through a DMC. But many other decisions may not require unblinded access. Some, including initiating a sample size re-assessment or updating a study's final statistical analysis plan (SAP), could be very problematic in terms of validly interpreting final analysis results if initiated by a party with access to unblinded interim results such as a DMC. In current practice and supported by prior regulatory guidance, such decisions are generally initiated by parties remaining blinded. Of course the DMC should be kept fully aware of any changes implemented in a trial, and should comment if they have any concerns. But for actions taken based upon blinded data, there are generally no confidentiality concerns, and sponsors can enlist any experts who can help arrive at the best decisions.

Establishing a qualified DMC when one was not previously felt to be needed can be challenging and time consuming during the pandemic. Attempting to ensure that DMC members have full understanding of all relevant background for the important tasks they will be assigned to, compared to trial personnel or steering committee members who will already have such perspective, could be risky. Thus, if an unblinded DMC is felt necessary to be established, given the challenges of identifying and implementing such a group quickly, an internal firewalled group might thus be considered as an option in some cases.

COVID-19 affects ongoing clinical trials in many different ways, which in turns affects many aspects of statistical inference, which are best described in the estimand framework laid out by \citet{ICHE9R1}. An estimand provides a precise description of the treatment effect reflecting the clinical question posed by the trial objective. It summarizes at a population-level what the outcomes would be in the same patients under different treatment conditions being compared. Central to the estimand framework introduced in \citet{ICHE9R1} are intercurrent events, which occur after treatment initiation that affect either the interpretation or the existence of the measurements associated with the clinical question of interest. Generally, the intercurrent events due to COVID-19 can be categorized into those that are more of an administrative or operational nature (e.g. treatment discontinuation due to drug supply issues), and those that are more directly related to the effect of COVID-19 on the health status of subjects (e.g. treatment discontinuation due to COVID-19 symptoms), see \citet{Akacha2020}. However, the additional intercurrent events are introducing ambiguity to the original research question and teams need to discuss how to account for them \citep{Akacha2017a, Akacha2017b,Qu2020}.

Care has to be taken when employing an adaptive design methodology to combine e.g. the information before and after the COVID-19 outbreak. When each stage of an adaptive design is based on a different estimand, the interpretability of the statistical inference may be hampered. If, for example, the pandemic markedly impacts the trial population after the outbreak because elderly and those with underlying conditions such as asthma, diabetes etc. are at higher risk and therefore excluded from the trial, then this would lead to different stagewise estimands (due to the different population attributes) and limit the overall trial interpretation. The situation is different in adaptive designs with a preplanned selection of a population at an interim analysis (as this does not change the estimand), when following the usual recommendations for an adequately planned trial (which includes the need to pre-specify the envisaged adaptation in the study protocol). Care has also to be taken if the pattern of intercurrent events is different before and after an interim analysis, in line with the usual recommendations to assess consistency across trial stages in an adaptive design. Generally speaking, as the definition of an adaptive design implies that we are considering a trial design, it needs to be aligned to the estimands that reflect the trial objectives according to \citet{ICHE9R1}.

The considerations in the previous paragraph are closely related to the trial homogeneity issues discussed in Section \ref{sec:heterogeneity}. One particular concern is the possible shift in the study population after the onset of the pandemic. At present we see a notable decline in hospital admissions for non COVID-19 related diseases. It can be assumed that patients with less severe problems tend to postpone a hospital stay for fear of an infection in the hospital or for not  putting stress on the already overloaded health system in some countries. Although standard trial procedures like randomization assures the validity of the statistical hypothesis test, it is unclear which population's treatment effect is actually being estimated.

The challenges imposed by the pandemic will lead to difficulties in meeting protocol-specified procedures in many instances, thus requiring the need to change aspects of ongoing trials. It is then important to be mindful about the fact that pre-specification of the study protocol and the SAP is the corner stone to avoid operational bias in any clinical trial. Although the \cite{ICHE9} guideline allows changing the SAP even shortly before unblinding a trial, this if often viewed as critical by stakeholders. Changing the characteristics of a trial based on unblinded trial data always requires appropriate measures to control the type I error rate whereas changes triggered by external data are often seen more lenient. In the case of changes to the conduct and/or analysis of a trial caused by the pandemic it is reasonable to assume that such changes are not triggered by the knowledge gained from the ongoing trial. Still, changes will have to be pre-specified and documented, as appropriate. It is recommended to pre-specify key analyses important to interpret the objectives of the trial in the statistical analysis plan, in particular analyses related to the inferential testing strategy. Therefore we suggest to consider first whether different analyses are needed for the primary or key secondary objectives. Other analyses that have a more exploratory character can be included in a separate exploratory analysis planning. If any impact is detected that warrant additional analyses in the clinical study report, then these can be added later.

After the lockdown measures will be eased in future, the medical practice may not return to the state before the onset of the pandemic. Social distancing measures may be kept in place and it is to be expected that the trend of, for example, fewer hospital admissions for minor cases will continue to some extent. Nevertheless, certain trials interrupted by the pandemic will be able to restart, albeit in a possibly changed environment. The  trial of the long acting contraceptive (Section \ref{sec:contra}) was largely unaffected by the onset of the pandemic. The START:REACTS trial (Section \ref{sec:startreacts}) had changed its endpoint to a PRO measure that can be observed remotely if a patient does not wish to come to the clinic. This trial can restart recruitment when elective surgeries will be again possible, albeit with the new endpoint as the original endpoint was not always measured during the lockdown measures. The ATALANTE 1 trial (Section \ref{sec:atalante}) was stopped for ethical reasons due to the study population being at high risk of COVID-19. As the trial did not proceed to its second stage, consultations with agencies have started to discuss the partial results in view of the clinical unmet need. Such discussion will be likely focus also on the loss of power even if first stage was promising. This begs the question, how promising the first stage results should have been to provide convincing evidence if a dilution of the treatment effect cannot be excluded a priori and the considerations in Section \ref{sec:resizing} of this paper may support such discussions. Lastly, the CAPE-Covid and the CAPE-Cod studies (Section \ref{sec:capecod}) are both addressing ICU patients with two kind of pneumonia. There is heterogeneity in disease and patients prognostic. For the moment, the  CAPE-Cod trial is temporarily stopped but is planned to restart next autumn. As the investigators had no choice than to embed a trial within the other, heterogeneity will need to be addressed at the end of the study in order to preserve both results.

\section{Discussion} \label{sec:discussion}
The COVID-19 pandemic has not only led to a surge a clinical research activities in developing treatments, diagnostics and vaccines to fight the pandemic, but also impacted in many ways on ongoing trials. Here we illustrated the negative effects the pandemic might have on trials by giving four examples from ongoing studies and describing the considerations and consequences in reaction to the pandemic. Furthermore, we focused here on the role of adaptive designs in mitigating the risks of the pandemic which might result in a large number of inconclusive or misleading trials. Aspects that are of particular importance here are type I error rate control and treatment-effect heterogeneity. When trials are affected, the question to stop the trial early or to continue the trial, possibly with modifications is of particular interest. Considering normally distributed outcomes we developed a range of strategies. We believe that these are transferable to other types of outcomes with only limited modifications.

\appendix
\section{Appendix}
\subsection{General joint distribution} \label{app:joint}
The joint distribution of $t_0$, $t_1$, and $t$ is given by
\begin{align}
\begin{pmatrix}t_0 \\ t_1 \\ t \end{pmatrix} \sim N &\left(
\begin{pmatrix}
\frac{\mu_{t0}-\mu_{c0}}{\sqrt{\frac{\sigma_{t0}^2}{n_{t0}} + \frac{\sigma_{c0}^2}{n_{c0}}}} \\
\frac{\mu_{t1}-\mu_{c1}}{\sqrt{\frac{\sigma_{t1}^2}{n_{t1}} + \frac{\sigma_{c1}^2}{n_{c1}}}} \\
\frac{\frac{n_{t0}\mu_{t0}+n_{t1}\mu_{t1}}{n_{t0}+n_{t1}}-\frac{n_{c0}\mu_{c0}+n_{c1}\mu_{c1}}{n_{c0}+n_{c1}}}{
\sqrt{\frac{n_{t0}\sigma_{t0}^2 + n_{t1}\sigma_{t1}^2}{\left(n_{t0}+n_{t1}\right)^2} + \frac{n_{c0}\sigma_{c0}^2 + n_{c1}\sigma_{c1}^2}{\left(n_{c0}+n_{c1}\right)^2}}}
\end{pmatrix}, \right.
\nonumber\\
&\quad\left.
\begin{pmatrix}
1 \\
0 & 1 \\
 \frac{\frac{\sigma_{t0}^2}{n_{t0}+n_{t1}} + \frac{\sigma_{c0}^2}{n_{c0}+n_{c1}}}{\sqrt{\left(\frac{\sigma_{t0}^2}{n_{t0}} + \frac{\sigma_{c0}^2}{n_{c0}}\right)
\left(\frac{n_{t0}\sigma_{t0}^2 + n_{t1}\sigma_{t1}^2}{\left(n_{t0}+n_{t1}\right)^2} + \frac{n_{c0}\sigma_{c0}^2 + n_{c1}\sigma_{c1}^2}{\left(n_{c0}+n_{c1}\right)^2}\right)}}
& \frac{\frac{\sigma_{t1}^2}{n_{t0}+n_{t1}} + \frac{\sigma_{c1}^2}{n_{c0}+n_{c1}}}{\sqrt{\left(\frac{\sigma_{t1}^2}{n_{t1}} + \frac{\sigma_{c1}^2}{n_{c1}}\right)
\left(\frac{n_{t0}\sigma_{t0}^2 + n_{t1}\sigma_{t1}^2}{\left(n_{t0}+n_{t1}\right)^2} + \frac{n_{c0}\sigma_{c0}^2 + n_{c1}\sigma_{c1}^2}{\left(n_{c0}+n_{c1}\right)^2}\right)}} & 1
\end{pmatrix}
 \right).
 \label{eqn:joint}
 \end{align}
Equation (\ref{eqn:joint}) can then be rewritten as
\begin{align}
\begin{pmatrix}t_0 \\ t_1 \\ t \end{pmatrix} \sim N &\left(
\begin{pmatrix}
\sqrt{\frac{N\tau}{r+1}} \frac{\delta}{\sqrt{\frac{\sigma_{t0}^2}{r} + \sigma_{c0}^2}} \\
\sqrt{\frac{N(1-\tau)}{r+1}} \frac{\left(1-\eta\right)\delta}{\sqrt{\frac{\psi_t \sigma_{t0}^2}{r} + \psi_c \sigma_{c0}^2}} \\
\sqrt{\frac{N}{r+1}} \frac{\delta\left(\tau+(1-\tau)\left(1-\eta\right)\right)}{\sqrt{\sigma_{t0}^2 \frac{\tau + (1-\tau)\psi_t}{r} + \sigma_{c0}^2  (\tau + (1-\tau)\psi_c)}}
\end{pmatrix},\right. \nonumber\\
&\quad\left.
 \begin{pmatrix}
1 \\
0 & 1 \\
\frac{\sqrt{\tau}\left(\frac{\sigma_{t0}^2}{r}+\sigma_{c0}^2\right)}{\sqrt{\left(\frac{\sigma_{t0}^2}{r}+\sigma_{c0}^2\right) \left(\sigma_{t0}^2 \frac{\tau+(1-\tau) \psi_t}{r} + \sigma_{c0}^2 \left(\tau+(1-\tau)\psi_c\right)\right)}}
& \frac{\sqrt{1-\tau}\left(\frac{\psi_t\sigma_{t0}^2}{r} + \psi_c \sigma_{c0}^2\right)}{\sqrt{\left(\frac{\psi_t\sigma_{t0}^2}{r} + \psi_c\sigma_{c0}^2\right)  \left(\sigma_{t0}^2\frac{\tau +(1-\tau)\psi_t}{r} + \sigma_{c0}^2 (\tau+(1-\tau) \psi_c)\right)}} & 1
\end{pmatrix}
\right).
\label{eqn:joint_simple_1}
\end{align}

\subsection{Sample Size Adjustment} \label{app:samplesize_adjustment}
Substituting $n_0/(n_0+\tilde{n}_1)$ with $\xi=n_0/(n_0+\tilde{n}_1)$, we can rewrite Equation (\ref{eqn:sample_size_adj}) as follows:
\begin{align}
&&\sqrt{\frac{(n_0+\tilde{n}_1)r}{(r+1)^2}} \cdot \frac{\frac{n_0}{n_0+\tilde{n}_1}+(1-\frac{n_0}{n_0+\tilde{n}_1})\left(1-\eta\right)}{\sqrt{\frac{n_0}{n_0+\tilde{n}_1}+ (1-\frac{n_0}{n_0+\tilde{n}_1})\psi}} \cdot \frac{\delta}{\sigma} - z_{1-\alpha} =& z_{1-\beta}  \nonumber\\
\Leftrightarrow && \sqrt{\frac{\frac{n_0}{\xi}r}{(r+1)^2}} \cdot \frac{\xi+(1-\xi)\left(1-\eta\right)}{\sqrt{\xi+ (1-\xi)\psi}} \cdot \frac{\delta}{\sigma} - z_{1-\alpha} =& z_{1-\beta} \nonumber\\
\Leftrightarrow && \frac{n_0}{\xi} \cdot \frac{\left(\xi+(1-\xi)\left(1-\eta\right)\right)^2}{\xi+ (1-\xi)\psi} =& \left(z_{1-\alpha} + z_{1-\beta}\right)^2 \frac{\sigma^2}{\delta^2} \frac{\left(r+1\right)^2}{r}
\end{align}
Replacing $n_0=N\tau$ and also noticing that the right-hand side of the equation also equals $N$, we obtain
\begin{align}
\Leftrightarrow && \frac{\tau}{\xi} \cdot \frac{\left(\xi+(1-\xi)\left(1-\eta\right)\right)^2}{\xi+ (1-\xi)\psi} =& 1 \nonumber\\
\Leftrightarrow && \xi^2\left(\tau\eta^2-1+\psi\right) + \xi \left(2\tau\eta\left(1-\eta\right)-\psi\right) + \tau\left(1-\eta\right)^2 =& 0.
\end{align}

Now, if $\tau\eta^2-1+\psi = 0$ and replacing $\psi=1-\tau\eta^2$, we get
\begin{align}
\xi =& \frac{\tau\left(1-\eta\right)^2}{1-\tau\eta\left(2-\eta\right)}.
\label{eqn:solution_xi_1}
\end{align}

For $\tau\eta^2-1+\psi \neq 0$, we obtain
\begin{align}
\xi = \frac{\psi - 2 \tau \eta (1-\eta) \pm \sqrt{\psi^2 - 4\tau(1-\eta)(\eta+\psi-1)}}{2\left(\psi - 1 + \tau \eta^2\right)}.
\label{eqn:solution_xi_2}
\end{align}
Re-substitution of $\xi$ finally yields
\begin{align}
\tilde{n}_1 =& N\tau \frac{1-\xi}{\xi} \nonumber\\
=& N\tau \frac{\psi - 2 + 2 \tau \eta \mp \sqrt{\psi^2 - 4\tau(1-\eta)(\eta+\psi-1)}}{\psi - 2 \tau \eta (1-\eta) \pm \sqrt{\psi^2 - 4\tau(1-\eta)(\eta+\psi-1)}}
\label{eqn:solution_n1_tilde}
\end{align}

As can be seen from Equations (\ref{eqn:solution_xi_2}) and (\ref{eqn:solution_n1_tilde}), $\xi$ and hence $\tilde{n}_1$ have two different solutions due to the square root. Evaluating both solutions for different values of $\tau$, $\eta$, and $\psi$ show that only the second solution ($+\sqrt{\phantom{x}}$ in the numerator and $-\sqrt{\phantom{x}}$ in the denominator lead to a positive number for the sample size.

\section*{Acknowledgment}
\noindent This work was inspired by discussions led by Lisa Hampson and Werner Brannath in the joint working group ``Adaptive Designs and Multiple Testing Procedures'' of the German Region (DR) and the Austro-Swiss Region (ROeS) of the International Biometric Society. The authors thank Andrea Schulze and Andy Metcalfe for their contributions to Examples 2.1 and 2.2, respectively. Furthermore, the authors are grateful to Gernot Wassmer for comments on the R code used in the R shiny app and to Werner Brannath for comments on an earlier version of this manuscript.

\section*{Conflict of Interest}
\noindent The authors have declared no conflict of interest.

\end{document}